\documentclass[%
 amsmath,amssymb,
 aps,prl,twocolumn
]{revtex4-2}
\usepackage{graphicx} 

\usepackage[utf8]{inputenc}
\usepackage{amstext,amssymb,amsfonts,bbm,amsthm}
\usepackage{mathtools}
\usepackage{hyperref}
\usepackage{graphicx}
\usepackage{bm}
\usepackage{xcolor}
\usepackage{booktabs} 
\usepackage{siunitx} 
\usepackage{pgfplots}
\usepackage{tikz}
\usepackage{svg}
\usepackage{cancel}
\usepackage{subfigure}
\usepackage{braket}
\usepackage{xr} 
\usepackage{hyperref}

\begin{document}

\title{Coarsening and universality on a growing surface}

\author{Robert J. H. Ross}
\email{robert.ross@oist.jp}
\author{Simone Pigolotti}
\email{simone.pigolotti@oist.jp}
\affiliation{Biological Complexity Unit, Okinawa Institute of Science and Technology, Onna, Okinawa 904-0495, Japan.}

\date{\today}

\begin{abstract}
We introduce a model in which cells belonging to two species proliferate with volume exclusion on an expanding surface. If the surface expands uniformly, we show that the domains formed by the two species present a critical behavior. We compute the critical exponents characterizing the decay of interfaces and the size distribution of domains using a mean-field theory. These mean-field exponents agree very accurately with those fitted in numerical simulations, suggesting that the theory is exact.
\end{abstract}

\maketitle

Living systems are characterized by an astonishing variety of patterns. These patterns are typically formed during development, and are orchestrated by complex dynamical processes. In most organisms, development is accompanied by growth, i.e., by a substantial increase in body size. The interplay between developmental dynamics and organisms growth can give rise to nontrivial physical phenomena. 

For example, domain growth can determine different outcomes in the paradigmatic example of Turing patterns \cite{Crampin1999,crampin2002pattern,krause2019influence} and in population models \cite{Ross2016b,Ross2017b}. More recently, our study of the arrangement of pigment cells on growing tissues has revealed that surface growth dramatically impacts disordered packing \cite{ross2025hyperdisordered}. In particular, random packing on a growing surface leads to a a critical-like behavior, with ordered regions without a characteristic size surrounded by more disordered regions. A critical-like behavior associated with homogeneous growth has also been observed in the context of morphogen gradient dynamics \cite{aguilar2018critical}.

These examples show that non-equilibrium models on growing domains display rich, unexpected physical behaviors. In contrast with traditional non-equilibrium lattice models  \cite{odor2004universality}, universality classes for growing systems are poorly explored.  Understanding this question could also shed light on a broader class of systems, such as those studied in the field of proliferating active matter \cite{Hallatschek2007, lavrentovich2014asymmetric, lavrentovich2015survival,dell2018growing,hallatschek2023proliferating}.  

In this Letter, we propose and study a model in which two species of cells proliferate on an expanding surface with volume exclusion. We call the model the ``growing voter model'' by way of analogy with the traditional voter model in non-equilibrium statistical physics \citep{dornic2001critical}. We find that coarsening in the growing voter model becomes critical for uniform growth. Nonuniform growth results in distinct phases. We predict the critical exponents associated with the decay of interfaces for uniform growth using a mean-field approach, and use this exponent to calculate the size distributions of species clusters. Simulation results suggest that the mean-field solution is exact.


\begin{figure}[h!]
\includegraphics[width = 0.45\textwidth]{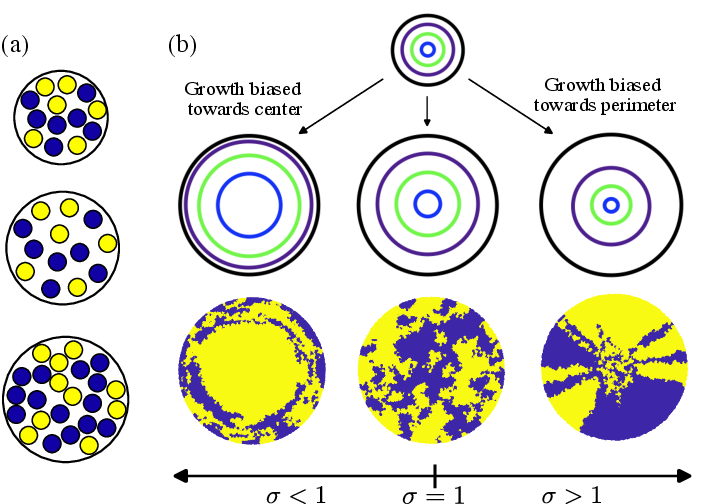}
\caption{Dynamics of the growing voter model. (a) Surface growth creates free space for proliferation of yellow (+) and blue (-) cells. (b) Contour lines associated with the displacement field $\lambda(\mu,\sigma,t)$. For $\sigma < 1$, this results in one cell type ultimately dominating the surface. When growth is uniform, $\sigma = 1$, neither cell type dominates and a seemingly critical behavior emerges. 
For $\sigma > 1$, the dynamics leads to the formation of sectors.}
\label{fig:fig1}
\end{figure}

We start by introducing the growing voter model. We consider a cell population consisting of two species of cells, that we denote by $+$ and $-$, placed on a growing circular surface of radius $R(t)$. We call $N_{+}(t)$ and $N_{-}(t)$ the numbers of $+$ and $-$ cells at time $t$, respectively.  The total number of cells is $N(t) = N_{+}(t) + N_{-}(t)$. Cells proliferate at a very large rate, identical for the two types. When a cell proliferates, it creates a new cell of the same type a distance $2c$ from its own center, where $c$ is a cell's radius, at an angle chosen uniformly at random, see Fig.~\ref{fig:fig1}a.  If the newly created cell overlaps with an existing cell, the proliferation event is aborted. The surface radially expands at a speed given by the displacement field $\lambda(r,\sigma,t) = (r/R(t))^{\sigma}$, where $r$ is the distance from the center of the surface. The parameter $\sigma>0$ tunes the stretching protocol. In particular, for $\sigma<1$ or $\sigma>1$, the surface grows more rapidly close to the center or boundary, respectively (see Fig.~\ref{fig:fig1}b). In the limiting case $\sigma=1$, the stretching rate is uniform.

The growing voter model generates different outcomes depending on how the surface grows, see Fig.~\ref{fig:fig1}b. In particular, for $\sigma<1$, one of the two types fixates at the center of the surface, while competition between the two types occurs at the perimeter. For $\sigma>1$, the two types occupy radial sectors, reminiscent of competing bacterial strains on agar plates \cite{Hallatschek2007}. In the case of uniform stretch ($\sigma=1$), the two species form domains without an apparent characteristic scale, suggesting a critical behavior. We shall focus on this case from now on.

\begin{figure}[h!]
\hspace{-0.75cm}
\includegraphics[width = 0.35\textwidth]{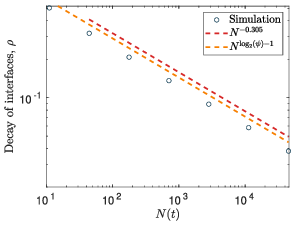}
\caption{Decay of interfaces for uniform surface growth, see Fig. \ref{fig:fig1} b.  An interface is defined as a pair of Voronoi neighbors that are of different colors. The fit of simulation data yields $N^{-0.305}$ with 95\% confidence intervals (-0.306, -0.304), red dashed line. For comparison, $\log_2(\psi)-1\approx 0.3057$.}
\label{fig:fig2}
\end{figure}

To characterize coarsening, we study the density of interfaces $\rho(t)$, defined as the number of Voronoi neighbor pairs that are of different colors divided by the total number of pairs. Simulations show that the density of interfaces decays as 
\begin{equation}\label{eq:decay_inter}
\rho \sim N(t)^{-\alpha}\, ,
\end{equation}
with $\alpha\approx 0.305$, see Fig. \ref{fig:fig2}. 

\begin{figure}
     \centering
     \includegraphics[width = 0.48\textwidth]{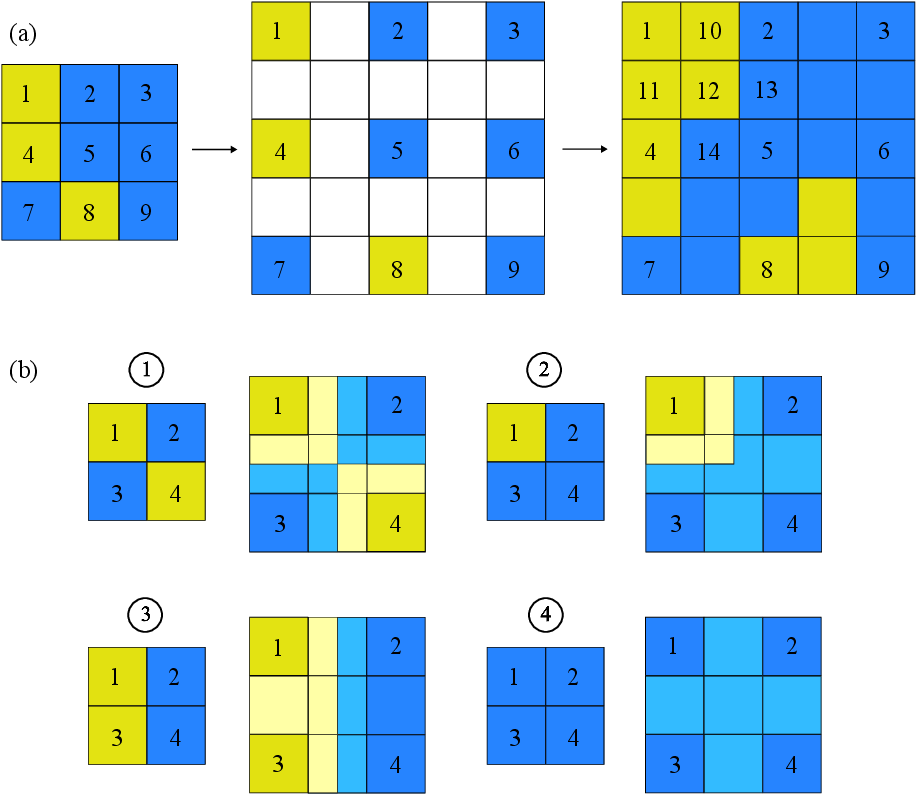}
     \caption{The two-dimensional lattice model. (a) At each time step the length of the domain in both the horizontal and vertical directions is doubled. The cell type of empty new sites are filled by selecting from their neighbors uniformly at random.  For instance, the new site 10 is filled by selecting from 1 or 2.  The new site 11 is filled by selecting from 1 or 4.  The new site 12 is filled by selecting from 1, 2, 4 or 5. (b) Doubling of elementary cells in the mean-field theory. The sites of a lighter shade indicate the new sites created by a growth event.}
    \label{fig:fig3}
\end{figure}

To rationalize this decay, we turn to a discrete time and discrete space analogous of the growing voter model. Each site of a two-dimensional lattice contains a $\pm 1$ cell. At each step, the sides of the lattice are doubled in length, and new empty sites are added to the lattice as illustrated in Fig. \ref{fig:fig3}a. Calling $n_0$ the number of sites on a side of the lattice at the initial time $k=0$, the system size at the $k^{th}$ step is $2^kn_0 \times 2^k n_0$. At each step, if a new empty site has two preexisting neighbors, one of these two neighbors is chosen uniformly at random to proliferate there. If a new site has no occupied neighbors, its color is uniformly and randomly chosen from one of the four preexisting diagonal sites. 

In the lattice model, we define an interface as two neighbors having different spin values who share an edge. The density of interfaces is the number of interfaces divided by the total number of neighbors ($n_0^2 2^{2k}$ at iteration $k$, assuming periodic boundary conditions).  We derive the decay rate of the density of interfaces by a mean-field assumption. We focus on a $2\times 2$ elementary cell. Assuming rotational and $Z_{2}$ symmetry, there are four possible types of elementary cells: 
\begin{align}
\raisebox{.5pt}{\textcircled{\raisebox{-.9pt} {1}}}=\begin{pmatrix} +1 & -1 \\ -1 & +1 \end{pmatrix}, \raisebox{.5pt}{\textcircled{\raisebox{-.9pt} {2}}}=\begin{pmatrix} +1 & -1 \\ +1 & -1 \end{pmatrix}, \\ \nonumber 
\raisebox{.5pt}{\textcircled{\raisebox{-.9pt} {3}}}=\begin{pmatrix} +1 & +1 \\ +1 & -1 \end{pmatrix}, \raisebox{.5pt}{\textcircled{\raisebox{-.9pt} {4}}}=\begin{pmatrix} +1 & +1 \\ +1 & +1 \end{pmatrix}. \nonumber
\end{align}
We calculate the proportions at which each of these configurations generate other ones at each step.  For example, configuration $\raisebox{.5pt}{\textcircled{\raisebox{-.9pt} {1}}}$ generates configurations $\underline{p} = \left(\raisebox{.5pt}{\textcircled{\raisebox{-.9pt} {1}}},\raisebox{.5pt}{\textcircled{\raisebox{-.9pt} {2}}},\raisebox{.5pt}{\textcircled{\raisebox{-.9pt} {3}}},\raisebox{.5pt}{\textcircled{\raisebox{-.9pt} {4}}}\right)$ in proportions $(1/8,1/4,1/2,1/8)$. We call $\underline{p}(k)$ the vector encoding the frequency of the four different elementary cell types in the system. The evolution of this vector is governed by a Markov chain
\begin{equation}\label{eq:markov}
    \underline{p}(k+1) = \hat{M}\underline{p}(k)\, ,
\end{equation}
where 
\begin{align}
\renewcommand{\arraystretch}{1.2}
    \hat{M} = \begin{pmatrix} \frac{1}{8} & 0 & \frac{1}{64} & 0 \\ \frac{1}{4} & \frac{1}{4} & \frac{5}{32} & 0 \\
    \frac{1}{2} & \frac{1}{2} & \frac{7}{16} & 0 \\ \frac{1}{8} & \frac{1}{4} & \frac{25}{64} & 1 \end{pmatrix}.
\end{align}
The leading eigenvalue of the matrix $\hat{M}$ is equal to 1, with an associated right eigenvector $(0,0,0,1)$. This eigenvector represents an absorbing state for Eq.~\eqref{eq:markov} in which only one of the two spins is present in the elementary cell. The second eigenvalue of $\hat{M}$ is equal to $(1 + \sqrt{5})/4$,  which implies that the interfaces asymptotically decay as
\begin{align}\label{eq:interface_decay}
\rho(2^{k}n_0 \times 2^{k}n_0) &\sim \left(\frac{1 + \psi}{4}\right)^{k}\, ,
\end{align}
where $\psi=(1+\sqrt{5})/2$ is the golden ratio.  To return to continuous time, we let $t \sim 2^{k}$, so that $k \sim \log_{2}(t) \sim \frac{1}{2}\log_{2}(N(t))$, and so we obtain
\begin{equation}\label{eq:interface_N}
\rho \sim N(t)^{-\alpha} \quad \mbox{with}\quad  \alpha=1-\log_{2}(\psi).
\end{equation}
The prediction of Eq.~\eqref{eq:interface_N} is in excellent agreement with numerical simulations of the off-lattice model, see Fig. \ref{fig:fig2}. This suggest that the solution is exact, and that the off-lattice and the lattice model belong to the same universality class. 

We briefly compare the scaling of interfaces between the growing voter model and the traditional non-growing voter model. In one dimension, interfaces are conserved in the growing voter model, and thus their density decays as $t^{-1}$ due to dilution. In contrast, in the one-dimensional, non-growing voter model, they present diffusive decay $t^{-1/2}$. In two dimensions, the scaling derived in this section contrasts with the logarithmic scaling found in the non-growing voter model \cite{dornic2001critical}. 
To understand this difference, we have derived the field theory for the growing voter model (see SI) and compared them with those for the traditional voter model \citep{dornic2005integration}. In one dimension, the field equation describing the growing voter model is equivalent to that of the voter model, but with a nonlinear time transformation in which time presents a critical slowdown \citep{Morris2014}. However, this equivalence does not hold in two dimensions.

\begin{figure}[h!]
\hspace{-0.5cm}
\includegraphics[width = 0.5\textwidth]{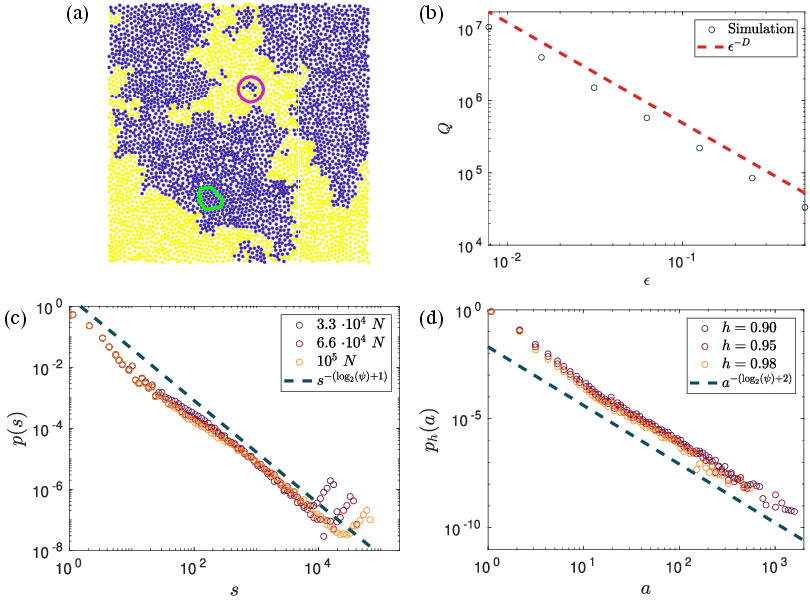}
\caption{(a) Examples of interfaces, clusters (magenta circle), and hexatic regions (green line) in the growing voter model. (b) Fractal dimension in the growing voter model. (c) Scaling of clusters in the growing voter model.  A cluster is defined as a Voronoi-connected component of cells of the same type. Each simulation replicate is initialized with 20 cells of randomly chosen type.  The total number of replicates for each system size is 500. A fit of simulation data for $10^{5}N$ yields $s^{-1.70}$ with 95\% confidence intervals (-1.73, -1.67).  (d) Scaling of hexatic regions in the growing voter model.  A hexatic region is defined as a Voronoi-connected component of cells that all have a hexatic order parameter greater than $h$. Each simulation replicate is initialized with 4 cells. The total number of replicates for each value of $h$ is 500. A fit of simulation data over the interval $s \in [10^1, 10^3]$ for $h = 0.9$ yields $s^{-2.68}$ with 95\% confidence intervals (-2.63,-2.74), $h = 0.95$ yield $s^{-2.66}$ with 95\% confidence intervals (-2.60,-2.72), and for $h = 0.98$ yields $s^{-2.71}$ with 95\% confidence intervals (-2.61,-2.80).}
\label{fig:fig4}
\end{figure}
\medskip

The decay of the interface density allows us to compute other observables in the growing voter model, see Fig.~\ref{fig:fig4}a. We begin by analyzing the fractal dimension $D$ of the interfaces, defined by
\begin{align}\label{eq:fd_definition}
Q = \epsilon^{-D},
\end{align}
where $\epsilon$ is a length scale and $Q$ is the number of segments of length $\epsilon$ necessary to measure the boundaries between the two species. In our case, the rescaling factor is analogous to growth: every time a growth event occurs, all distances within the system are halved, which is equivalent to dividing $\epsilon$ by two. Given this, we write Eq.~\eqref{eq:fd_definition} for our system as
\begin{align}
(4^{k}n^2_{0})^{1-\alpha} \sim \left(\frac{1}{2^{k}}\right)^{-D}\, .
\end{align}
which implies
\begin{align}\label{eq:fractal_dimension}
D = 2(1-\alpha)\, .
\end{align}
Substituting the value of $\alpha$ given by Eq.~\eqref{eq:interface_N}, we obtain $D= \log_{2}(1 + \psi) = 1.388483...$, which is twice the fractal dimension of the asymmetric Cantor set ($\log_{2} (\psi)$). This prediction is in excellent agreement with numerical simulations, see Fig.~\ref{fig:fig4}b.

We now analyze the distribution of cluster sizes. We define as a cluster a connected set of Voronoi neighbors having the same spin value (color). For uniform expansion, the distribution of cluster sizes for large $N$ appears to decay as a power law:
\begin{equation}\label{eq:clust_critical}
p(a) = \frac{n(s)}{\Tilde{n}} \sim s^{-\beta}\, ,
\end{equation}
where $n(s)$ is the number of clusters of size $s$ and $\Tilde{n}$ is the total number of clusters.

To compute the exponent $\beta$, we assume that  clusters grow  proportionally to their area. This assumption implies
\begin{align}\label{eq:PDE_solution}
s(N;N_{0}) = \left(\frac{N}{N_{0}}\right)\, ,
\end{align}
where $s(N;N_{0})$ is the size of a cluster at system size $N$ that was seeded at system size $N_{0}$. We now further assume that clusters are seeded proportionally to the number of interfaces in the system. This assumption implies $d\Tilde{n}/dN\ \sim N^{-\alpha}$, see Eq.~\eqref{eq:decay_inter}, and therefore
\begin{align}\label{eq:n_tilde}
\Tilde{n} \sim N^{1-\alpha}\, .
\end{align}
We now rewrite Eq. \eqref{eq:PDE_solution} in terms of numbers of clusters, expressing only the dependence on $\Tilde{n}_0$:
\begin{equation}\label{eq:cluster_age}
s(\Tilde{n}_{0}) \sim \Tilde{n}_{0}^{-1/(1 - \alpha)}\, .
\end{equation}
Picking a random cluster amounts to uniformly sampling $\Tilde{n}_0$, so that $p(s)\propto d\Tilde{n}_0/ds$. Using Eq.~\eqref{eq:clust_critical} and Eq.~\eqref{eq:cluster_age} we therefore obtain
\begin{align}
\beta = 2 - \alpha = 1.6942...
\end{align}
This prediction is in excellent agreement with the simulation results, see Fig.~\ref{fig:fig4}c.

Ignoring species identity, cell configurations in the growing voter model are characterized by ordered regions surrounded by disordered ones, see Fig.~\ref{fig:fig4}a and \cite{ross2025hyperdisordered}. 
The behavior of these hexatic regions is highly fluctuating, unlike the clusters, in which hexatic regions fragment and aggregate between larger and smaller ordered regions (see SM video 1).  To measure these regions, we introduce the hexatic order parameter
$$h_{j} ={\frac {1}{N_{j}}}\sum _{k=1}^{N_{j}}e^{i6\theta _{jk}},$$
where $j$ is an index representing a given cell, $N_{j}$ is the number of nearest neighbors of cell $j$ computed via Delauney triangulation, and $\theta_{jk}$ is the angle between the vector pointing from cell $j$ to $k$ and the $x$ axis.  We designate a cell as ordered if its hexatic order parameter is larger than a threshold $h$, and a hexatic region is then defined as a Voronoi-connected component of cells that all have a hexatic order parameter greater than $h$. For large times, the expected density of ordered regions approaches a finite constant (that depends on $h$). Simulations also indicate that new hexatic regions are seeded at a constant rate.

We now compute the distribution of hexatic region sizes. For uniform expansion, the distribution of hexatic region sizes for large $N$ also appears to decay as a power law:
\begin{equation}\label{eq:hex_critical}
p_h(a) = \frac{n_h(a)}{\Tilde{n}_{h}} \sim s^{-\beta_{h}}\, ,
\end{equation}
where $n_h(a)$ is the number of hexatic regions of size $a$ and $\Tilde{n}_{h}$ is the total number of hexatic regions. As in the case of clusters, we assume hexatic regions to grow proportionally to their area, see Eq. \eqref{eq:PDE_solution}. Furthermore, we assume within a cluster the distribution describing the size of hexatic regions evolves independently, so that within a cluster $p_h(a|s) \sim a^{-2}$. As such, the cluster distribution describes how the entire cell population `fractures' into independently evolving sections as the system grows. Simulations support this assumption (see SI). To finish we compute the marginal probability distribution of $p(s_{h})$ over the cluster size distribution
\begin{equation}\label{eq:hexregion_integral}
p_h(a) \sim \int_{s=a}^{\infty}p(s)p_h(a|s)ds \sim  a^{-2}\int_{s=a}^{\infty}p(s)ds ,
\end{equation}
and so
\begin{equation}\label{eq:hexregion_dist}
\beta_{h} = 3 - \alpha = 2.6942...
\end{equation}
This prediction is in excellent agreement with the simulation results of the hexatic region scaling, see Fig. \ref{fig:fig4}d.

In conclusion, in this Letter we have proposed the growing voter model as a paradigm for the statistical physics of growing systems. We have identified the case of uniform stretch as a critical point for this system, and characterized the associated critical exponents. In doing so, we have demonstrated a connection to another model of circle packing on a growing two-dimensional surface \citep{ross2025hyperdisordered}. 

We have solved a lattice version of the growing voter model using mean field. The predicted critical exponents of this mean field theory match very well the simulations of the growing voter model. This agreement suggests that the solution is exact, and that the lattice and off-lattice models belong to the same universality class. The success of the mean field theory could be due to the fact that growth causes lattice sites to behave independently of their neighbors. A similar effect has been studied in the evolution of diffusive processes at the perimeter of growing circles \citep{Hallatschek2007, takeuchi2018appetizer}, and also reported in dynamical processes situated on growing networks \citep{ross2019compressibility}. At a larger scale, it is similar to non-gravitationally coupled massive structures in the universe, and their diminishing interactions as they move further apart due to cosmic inflation \citep{baumann2018tasi}. 

It is interesting to ask to what extent surface growth is analogous to proliferating active matter.  For instance, bacterial colonies can exhibit radial expansion akin to if they were situated on a growing surface, but instead caused by cell proliferation \citep{dell2018growing, you2018geometry, Tjhung2020}. Similarly, microfluidic devices allow bacteria to grow in a manner equivalent to one-dimensional expansion \citep{koldaeva2022population}.
Ultimately, it may be the case that active matter on a growing surface, and active matter that expands due to proliferation, are governed by similar physical principles.

\begin{acknowledgments}
We thank Samuel Cure and Sam Reiter for useful discussions.
\end{acknowledgments}

\bibliography{References} 


\end{document}